\documentclass[conference,a4paper]{IEEEtran}
\IEEEoverridecommandlockouts

\usepackage{graphicx}
\usepackage{amsmath}
\usepackage{amssymb}
\usepackage{color}
\usepackage{multirow}
\usepackage{tabularx}
\usepackage{balance}
\usepackage[numbers]{natbib}
\usepackage{url}

\usepackage{algorithm}
\usepackage{algpseudocode}

\algnewcommand\algorithmicinput{\textbf{Input:}}
\algnewcommand\INPUT{\item[\algorithmicinput]}
\algnewcommand\algorithmicoutput{\textbf{Output:}}
\algnewcommand\OUTPUT{\item[\algorithmicoutput]}

\usepackage{mathtools}

\newcommand{\Fig}[1]{Fig.~\textup{\ref{#1}}}
\graphicspath{{figures/}}
\usepackage{subcaption}

\newtheorem{remark}{Remark}
\newtheorem{example}{Example}

\newcommand{\weight}[1]{\mathop{wt}\left(#1\right)}

\begin{document}

\title{Design and Decoding of Polar Codes for the Gaussian Multiple Access Channel\thanks{The research was carried out at Skoltech and supported by the Russian Science Foundation (project no. 18-19-00673).}}

\author{
  \IEEEauthorblockN{Evgeny Marshakov\IEEEauthorrefmark{1}, Gleb Balitskiy \IEEEauthorrefmark{1}\IEEEauthorrefmark{2}, Kirill Andreev\IEEEauthorrefmark{1} and Alexey Frolov\IEEEauthorrefmark{1}}
    
 \IEEEauthorblockA{\small \IEEEauthorrefmark{1} Skolkovo Institute of Science and Technology, Moscow, Russia
    }
\IEEEauthorblockA{\small \IEEEauthorrefmark{2} Moscow Institute of Physics and Technology \\
    Moscow, Russia
    }

  {evgeny.marshakov@skoltech.ru, gleb.balitskiy@skoltech.ru, k.andreev@skoltech.ru, al.frolov@skoltech.ru}
      
}

\maketitle
\begin{abstract}
    Massive machine-type communications (mMTC) is one of the key application scenarios for future 5G networks. Non-orthogonal multiple access (NOMA) is a promising technique for the use in mMTC scenario. In this paper, we investigate NOMA schemes based on polar codes. We compare two possible decoding techniques: joint successive cancellation algorithm and joint iterative algorithm. In order to optimize the codes (choose frozen bits) we propose a special and efficient design algorithm. We investigate the performance of the resulting scheme in the Gaussian multiple access channel (GMAC) by means of simulations. The scheme is shown to outperform LDPC based solution by approximately $1$~dB and to be close to the achievability bound for GMAC.
\end{abstract}

\section{Introduction}

The main goal of existing wireless networks is to provide the highest possible spectral efficiency and the best possible data rate for human users. But, machine-type communications (MTC) will become a strong challenge for next generation wireless networks. Traffic patterns for MTC completely differ from human-generated traffic and can be characterized by the following features: (a) a huge number of autonomous devices connected to one access point, (b) low energy consumption is a vital requirement, (c) short data packets and (d) low traffic intensity generated by single device. 3GPP has proposed multiple candidate solutions for massive MTC (mMTC). The main candidates are multi-user shared access (MUSA, \cite{yuan2016non}), sparse coded multiple access (SCMA, \cite{nikopour2013sparse}) and resource shared multiple access (RSMA, \cite{3gpp.R1-164688, 3gpp.R1-164689}), but the lack of implementation details does not allow to select the most preferable solution. At the same time, we mention that no one of 3GPP solutions is based on polar codes \cite{Arikan} despite the fact that these codes in combination with Tal--Vardy list decoder \cite{TalVardyList2015} are extremely good for short code lengths and low code rates. In this paper, we fill this gap.

Polar codes \cite{Arikan} are the first class of error-correcting codes which is proved to achieve the capacity of any binary memoryless symmetric channel with a low-complexity encoding and
decoding procedures. However, it appeared to be a challenging problem to construct (optimize) such codes for the finite blocklength regime. This question was addressed in \cite{TalVardyConstr2013, MoriTanaka2009, Trifonov2012}. These methods as well as Tal--Vardy list decoder \cite{TalVardyList2015} allowed to significantly improve the practical performance of such codes. As a result, these codes were selected as a coding scheme for the control channel of the enhanced mobile broadband (eMBB) \cite{3gpp.finrep, 3gpp.R1-1611109}.

In \cite{Telatar2user, Onay2013} polar codes were proved to achieve the full admissible capacity region of the two-user binary input MAC. In \cite{TelatarMuser} the results were generalized for the $K$-user case. At the same time, there are no efficient decoding and optimization methods for the case of finite blocklength. In this paper, we address this question and investigate the practical performance of polar codes in $K$-user MAC. 

Our contribution is as follows. We compare two possible decoding techniques: joint successive cancellation algorithm and joint iterative algorithm. In order to optimize the codes (choose frozen bits) we propose a special and efficient design algorithm. We investigate the performance of the resulting scheme in the Gaussian multiple access channel (GMAC) by means of simulations. The scheme is shown to outperform LDPC based solution by approximately $1$~dB and to be close to the achievability bound for GMAC.

\section{Preliminaries}

\subsection{Polar Codes}
Let us consider the Arikan's kernel
\[
G_2 \triangleq \begin{bmatrix} 1 & 0 \\
1 & 1
\end{bmatrix},
\]
then the \textit{polar transform} of size $N = 2^n$ is defined as follows
\[
G_N \triangleq B_N G_2^{\otimes n},
\]
where $\otimes$ is the Kronecker power and $B_N$ is called a \textit{shuffle reverse} operator (see \cite{Arikan}).

In order to construct an $(N, k)$ polar coset code let us denote the set of frozen positions by $\mathcal{F}$, $|\mathcal{F}| = N - k$. By $\mathbf{u}_\mathcal{F}$ we denoted the projection of the vector $\mathbf{u}$ to positions in $\mathcal{F}$. For now, we can define a \textit{polar coset code} $\mathcal{C}$ as follows
\[
\mathcal{C}(N, k, \mathcal{F}, \mathbf{f}) = \left\{ \mathbf{c} = \mathbf{u} G_N \:\: | \:\: \mathbf{u} \in \{0,1\}^N, \:\: \mathbf{u}_\mathcal{F} = \mathbf{f} \right\}.
\]


\subsection{System model}

Let us describe the system model. There are $K$ active users in the system. Communication proceeds in a frame-synchronized fashion. The length of each frame is $N$ and coincides with the codeword length. Each user has $k$ bits to transmit during a frame. All users have equal powers and code rates.

Let us describe the channel model
\begin{equation*}
\mathbf{y} = \sum_{i=1}^{K} \mathbf{x}_i + \mathbf{z},   
\end{equation*}
where $\mathbf{x}_i \in \mathbb{R}^n$ is a codeword transmitted by the $i$-th user and $\mathbf{z} \sim \mathcal{N}(\mathbf{0}, \mathbf{I})$ is an additive white Gaussian noise (AWGN).

We note, that non-asymptotic achievability and converse bounds for this channel were derived in \cite{polyanskiy2017perspective}. We note, that these bounds were proved for the case of the same codebook and decoding up to permutation, but can be easily changed for the use in different codebook case. In what follows we compare the performance of our codes to these bounds.

In our system the users utilize \textit{different} polar coset codes $\mathcal{C}_i(N, k, \mathcal{F}_i, \mathbf{f}_i)$, $i=1,\ldots, K$. Lets us consider the \mbox{$i$-th} user. In order to send the information word $\mathbf{u}_i$ the user first encodes it with the code $\mathcal{C}_i(N, k, \mathcal{F}_i, \mathbf{f}_i)$ and obtain a codeword $\mathbf{c}_i$. Then the user performs BPSK modulation or equivalently 
\[
\mathbf{x}_i = \tau(\mathbf{c}_i), \quad \tau(\mathbf{c}_i) = (\tau(c_{i,1}), \ldots, \tau(c_{i,N})),
\]
where $\tau:\{0, 1\} \rightarrow \{\sqrt{P}, -\sqrt{P}\}$.

The probability of error (per user) is defined as follows
\begin{equation}
\label{eq:p_e}
P_e = \frac{1}{K} \sum\limits_{i=1}^{K} \Pr(\mathbf{u}_i \ne \hat{\mathbf{u}}_i),
\end{equation}
where $\hat{\mathbf{u}}_i$ is the estimate of $\mathbf{u}_i$ provided by the decoder.

As energy efficiency is of critical importance for mMTC scenario we focus on optimization of the required energy per bit ($E_b/N_0$). Recall, that it is calculated as follows
\[
E_b/N_0 = \frac{N P}{2 k}.
\]

\section{Decoding algorithms}

\subsection{Joint Successive Cancellation Decoding}

\begin{figure}[t]
    \centering
    \includegraphics[width=0.8\columnwidth]{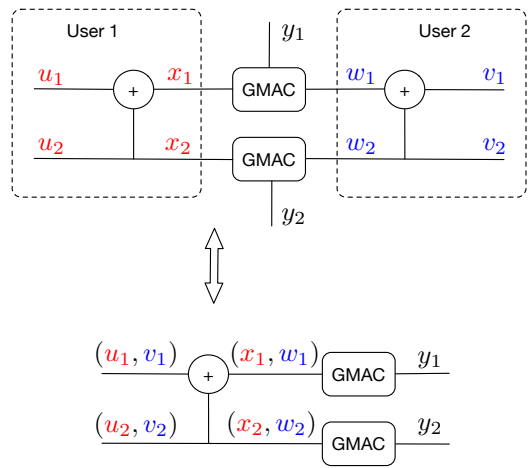}
    \caption{Representation as a polar code over $\mathbb{Z}_2^K$ for $K=2$.}
    \label{fig:PolarRepr}
\end{figure}

Let us first explain the main idea for the toy example with $N=2$ (see \Fig{fig:PolarRepr}). We see that instead of working with bits of different users and several polar codes we can work with a single polar code over $\mathbb{Z}_2^K$. In our example we will first decode a bit configuration (tuple) $(u_1, v_1)$ -- first bits of users and then a tuple $(u_2, v_2)$ -- second bits of users. We assume the decoder to work with tuple distributions rather than with probabilities of single bits.

The input of the decoder is the vector $\mathbf{P} = ({\mu}_1,\ldots,{\mu}_N)$ of length $N$ consisting of a priory probability mass functions (pmf) ${\mu}_i \in [0, 1]^{2^K}$, $i = 1, \ldots, N$. Let us show how to initialize the $k$-th pmf. Recall, that the channel output is a vector $\mathbf{y}$ and consider its $k$-th component $y_k$ and let $g = (b_1, \ldots, b_K) \in \mathbb{Z}_2^K$.

\begin{flalign}
&{\mu}_k(g) = \Pr[g = (b_1, \ldots, b_K) | y_k] \nonumber \\
&\propto \exp \left\{-\frac{(y_k - \sum_{i=1}^K \tau(b_i))^2}{2\sigma^2}\right\}, \label{eq:initialization}
\end{flalign}
recall, that the noise variance $\sigma^2 = 1$ in our case.

Let us first consider the decoding of the basic block shown in \Fig{fig:PolarRepr}. Let us assume, that we are given two a priory pmfs $\mathbf{\mu}_1$ and $\mathbf{\mu}_2$. Let us describe the operations. 

We start with decoding of the tuple corresponding to the first bits of the users ($(u_1, v_1)$ in our example). In order to do this, we need to calculate the distribution of the sum of two random variables over $\mathbb{Z}_2^K$. In what follows we refer to this operation as the \textit{check-node operation (cnop)}. Clear, that this can be done by means of convolution, i.e. $\hat{\mu}_1 = \mu_1 \ast \mu_2$. As we are working in the abelian group $\mathbb{Z}_2^K$, so there exists a Fourier transform (FT) $\mathcal F$. In what follows in order to perform a convolution we use the FFT-based technique proposed in \cite{Declercq} the case of LDPC codes over Abelian groups. Thus, the final rule is as follows
\begin{equation*}
\hat{\mu}_1 \propto \mathcal F^{-1}\left(\mathcal F(\mu_1) \odot \mathcal F(\mu_2)\right),
\end{equation*}
where $\odot$ denotes the element-wise multiplication.

After we calculated the pmf $\hat{\mu}_1$ we can make a hard decision $\hat{g}_1$ taking into account the values of frozen bits in this position. 

After $\hat{g}_1$ is found we proceed with \textit{variable-node operation (vnop)}. The rule is as follows
\begin{equation*}
\hat{\mu}_2(g) \propto \mu_1(g + \hat{g}_1) \mu_2(g)\ \ \ \forall g\in \mathbb Z_2^K.
\end{equation*}


The final joint successive cancellation (JSC) decoding algorithm utilizes $cnop$ and $vnop$ functions in a recursive manner. Please see Algorithm~\ref{alg:jsc} for full description.

\begin{algorithm} 
\caption{Joint Successive Cancellation Decoding (JSC)}
\label{alg:jsc}
\begin{algorithmic}[1]
\INPUT{$N$ -- code length, $K$ -- number of users, $\mathbf{F} \in \{0, 1, \text{inf}\}^{K \times N}$ -- matrix of frozen bits, $\mathbf{y} \in \mathbb{R}^N$ -- received signal.}
\State Initialize $\mathbf{P} = ({\mu}_1,\ldots,{\mu}_N)$ according to \eqref{eq:initialization}.
\Function{PolarDecode}{$\mathbf{P}$, $\mathbf{F}$}
\If{$ \mathop{len}(P) = 1$} 
\State $u, x = \mathop{decision}(\mathbf{P}, \mathbf{F})$ 
\Comment{Make decision based on probabilities and the matrix of frozen bits}
\Else
\State $\mathbf{P}_o = ({\mu}_1, {\mu}_3, \ldots)$, $\mathbf{P}_e = ({\mu}_2, {\mu}_4, \ldots)$ 
\State $\mathbf{P}_{1} = \mathop{cnop}(\mathbf{P}_e, \mathbf{P}_o)$
\State $\mathbf{u}_1, \mathbf{x}_1 = \mathop{PolarDecode}(\mathbf{P}_{1}, \mathbf{F})$ 
\State $\mathbf{P}_{2} = \mathop{vnop}(\mathop{cnop}(\mathbf{x}_1, \mathbf{P}_o), \mathbf{P}_e)$
\State $\mathbf{u}_2, \mathbf{x}_2 = \mathop{PolarDecode}(\mathbf{P}_{2}, \mathbf{F})$
\State $\mathbf{u} = \mathop{concat}(\mathbf{u}_1, \mathbf{u}_2)$
\State $\mathbf{x} = \mathop{merge}(\mathbf{x}_1, \mathbf{x}_2)$
\EndIf
\EndFunction
\OUTPUT{$\mathbf{u}$, $\mathbf{x}$}
\end{algorithmic}
\end{algorithm}

\begin{remark}
It is worth noting that we can easily improve the decoding procedure by using list decoding method \cite{TalVardyList2015}. It means we will take into account not only the most probable path (the path consists of tuples in our case) but $L$ different paths with the highest metric. 
\end{remark}

\subsection{Iterative Decoding}
\begin{figure}[t]
    \centering
    \includegraphics[width=0.8\columnwidth]{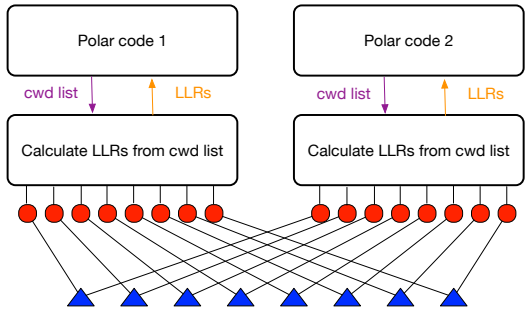}
    \caption{Iterative decoding scheme for $K=2$.}
    \label{fig:IterScheme}
\end{figure}

Now let us describe the iterative decoding algorithm. The aim of this decoder is to update the log-likelihood ratios (LLR) for every bit that has been transmitted by every user. During an iteration the algorithm selects the next user from the list in a round robin manner, fixes the remaining LLRs and updates the LLR vector only for the user under consideration. Every iteration consists of a message passing algorithm on a graph shown in Figure~\ref{fig:IterScheme}. This graph has the following nodes: a) the polar list decoder node, which uses the LLR values as inputs and generates $L$ candidate codewords as the output, b) LLR evaluation nodes (circles), which perform the per-bit LLR evaluation given the input candidate codewords list~\citep{Pyndiah}, and c) the functional nodes (represented by triangles) performing the per-user LLR update given the LLR vectors for every user and the received signal vector $\mathbf{y}$.

As soon as the polar list decoding is a well-known procedure~\citep{TalVardyList2015} as well as the method of constructing the LLR vector from the candidate codeword list~\citep{Pyndiah}, one need to describe in details the message passing procedure to and from the functional nodes.
\begin{algorithm} 
\caption{Iterative Decoding}
\label{alg:polar_iterative}
\begin{algorithmic}[1]
\INPUT{$N$ - code length, $K$ - number of users, $\mathbf{F}\in \{0, 1, inf\}^{K\times N}$ - matrix of frozen bits, $\mathbf{y}\in \mathbb{R}^N$ - received signal.}
\State initialize the LLR values of variable nodes for each user code with zero values assuming equal probability for $\sqrt{P}$ and $-\sqrt{P}$ values
\For {$i = 1, \ldots, K\times I$} \Comment perform $I$ iterations
\State {$u = mod(I, K)$} \Comment round robin user selection
\State Update LLR vector for given user assuming all other users have fixed LLRs~eq.~\eqref{eq:update_func_nodes} \Comment from functional nodes to polar decoder
\State Perform single user list polar decoding~\citep{TalVardyList2015} given the input LLR vector \Comment corresponds to orange arrow on Figure~\ref{fig:IterScheme}
\State Derive output LLR vector given the decoded candidate list \Comment corresponds to magenta arrow on Figure~\ref{fig:IterScheme}
\EndFor
\State Make decisions given the output LLR vector for every user
\OUTPUT{u, x}
\end{algorithmic}
\end{algorithm}
Every functional node corresponds to a single channel use. As mentioned above, every user's LLR vector is updated under fixed LLR vectors for all other users. For convenience let us consider some arbitrary functional node (its index is omitted for convenience) and the first user. The goal of the functional node is to marginalize out the uncertainty about the signal transmitted by users $j=2,\ldots,K$
\begin{equation}
L(x_1) = \log \left(\frac{
    \sum\limits_{x_1 = +\sqrt{P}, x_2, \ldots x_K} p\left(y \bigg| \sum\limits_{j=1}^{K} x_j\right)\prod\limits_{j=2}^{K} \Pr(x_j)
}
{
    \sum\limits_{x_1 = -\sqrt{P}, x_2, \ldots x_K} p\left(y \bigg| \sum\limits_{j=1}^{K} x_j\right)\prod\limits_{j=2}^{K} \Pr(x_j)
}
\right),
\label{eq:update_func_nodes}
\end{equation}
where the numerator corresponds to the total probability that user $1$ has transmitted the signal $x_1=+\sqrt{P}$ and the denominator -- that $x_1=-\sqrt{P}$ has been transmitted (subscript corresponds to user number) and $L(x_1)$ is the output LLR for the first user. The probability $p(y|a) = \frac{1}{\sqrt{\pi}}\exp{\left(-(y-a)^2\right)}$ corresponds to AWGN channel assumption. Full algorithm description is presented in~Algorithm~\ref{alg:polar_iterative}.

\section{Design of Polar Codes for GMAC}

\begin{figure}[t]
    \centering
    \includegraphics[width=0.9\columnwidth]{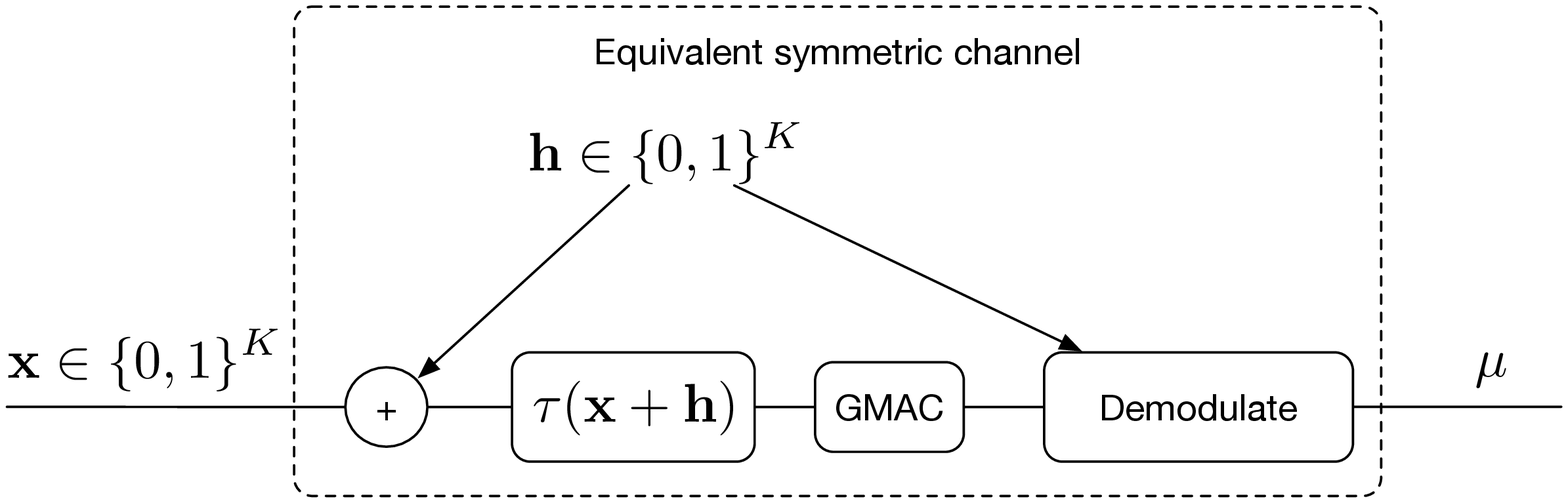}
    \caption{Equivalent symmetric channel}
    \label{fig:SGMAC}
\end{figure}

In this section, we propose a method to optimize polar codes for the use in $K$-user GMAC. First of all, let us mention the fact, that GMAC is not a symmetric channel. To see this fact let us consider $K=2$ and a noiseless case. We see, that the tuples $(0,1)$ and $(1,0)$ will lead to the channel output $0$ and it is not possible to distinguish in between this two hypotheses given $y = 0$. At the same time $(0,0)$ and $(1,1)$ will lead to $y = 2$ and $y = -2$ and the decoder can easily find the transmitted tuple. Thus, the zero codeword assumption (so popular in the single user case) does not work in our case. In order to construct the codes we apply the approach of \cite{SGMAC} and ``symmetrize'' the channel (see \Fig{fig:SGMAC}). The main idea is in adding and subtracting (during demodulation process (\ref{eq:initialization})) of a random element $\mathbf{h}$ distributes uniformly on $\mathbb{Z}_2^K$ (different $\mathbf{h}$ for each channel use). 

It is easy to see that the resulting channel (see \Fig{fig:SGMAC}) is symmetric. In what follows we refer to it as sGMAC and construct the codes for it. For this channel, we can use a zero codeword assumption.

Initially, we supposed to write density evolution rules for our case. The idea is very similar to density evolution for non-binary LDPC codes \cite{NBDE_LDPC}. To be precise we mean Gaussian approximation, i.e. pdfs of the messages are approximated with use of multidimensional (the dimension is $2^K$) Gaussian mixtures. But we found that this procedure requires much more computational resources in comparison to simple Monte-Carlo simulation to determine good subchannels. The problem is in $\mathop{cnop}$ operation. It is rather difficult and requires sampling and fitting operations.

Finally, we found that the major problem for the decoder is non-unique decoding rather than the noise and propose a construction method for the noiseless adder MAC, which works fine for sGMAC also. Let us briefly define the method. We suppose that zero tuples are being transmitted through symmetric noiseless MAC (see \Fig{fig:SGMAC}). First of all we need to calculate the initial pmf, which goes to the decoder. It can be done as follows 
\[
{\mu}_0(\mathbf{x}) = \frac{1}{2^K}\sum\limits_{\mathbf{h}: \:\: \weight{\mathbf{h} + \mathbf{x}} = \weight{\mathbf{h}}} \frac{1}{\binom{K}{\weight{\mathbf{h}}}},
\]
by $\weight{\cdot}$ we mean the Hamming weight, i.e. a number of non-zero elements in a vector.

\begin{example}
Let $K = 2$ and assume, that the users send a tuple $(0, 0)$. Then consider $4$ cases of $\mathbf{h}$ and calculate $\mu_0$ for this case:
\begin{enumerate}
    \item $\mathbf{h} = (0,0), \quad \mu = \begin{bmatrix} 1 & 0 & 0 & 0 \end{bmatrix}$;
    \item $\mathbf{h} = (0,1), \quad \mu = \begin{bmatrix} 1/2 & 0 & 0 & 1/2 \end{bmatrix}$;
    \item $\mathbf{h} = (1,0), \quad \mu = \begin{bmatrix} 1/2 & 0 & 0 & 1/2 \end{bmatrix}$;
    \item $\mathbf{h} = (1,1), \quad \mu = \begin{bmatrix} 1 & 0 & 0 & 0 \end{bmatrix}$;
\end{enumerate}

Thus, the resulting initial distribution (averaged over $h$) is $\mu_0 = \begin{bmatrix} 3/4 & 0 & 0 & 1/4 \end{bmatrix}$. The elements of $\mu_0$ are indices by the tuples in lexicographic order.
\end{example}

At each of $\nu = 0, \ldots, n-1$, with $n = \log_2 N$ steps we construct $2$ new pmfs: $\mu_{\nu+1}^-$ and $\mu_{\nu+1}^+$
$$
\mu_{\nu+1}^- = \mathop{cnop} (\mu_{\nu}, \mu_{\nu}),\quad \mu_{\nu+1}^+ = \mathop{vnop} (\mu_{\nu}, \mu_{\nu}),
$$
where $\mu_{\nu+1}^{(2i-1)} = \mu_{\nu+1}^{-,(i)}$, $\mu_{\nu+1}^{(2i)} = \mu_{\nu+1}^{+, (i)}$.
To choose the subchannels we compare $\mu_{n}(\mathbf{0})$ values on the $n$-th step.

\section{Numerical Results and Experiments}
We conducted a series of experiments of proposed algorithms and compared the results with GMAC random coding bound \cite{polyanskiy2017perspective} and with PEXIT optimized LDPC code ($15$ inner and $15$ outer iterations) proposed in \cite{10.1007/978-3-030-01168-0_15}. 

Let us describe how we constructed polar codes for our experiments. In order to choose the frozen positions we utilized the proposed design procedure. We have selected the common set of frozen tuples for all users and the values of frozen bits have been selected at random for different users because the same frozen values lead to poor performance.

\begin{figure}[t]
    \centering
    \includegraphics[width=0.8\columnwidth]{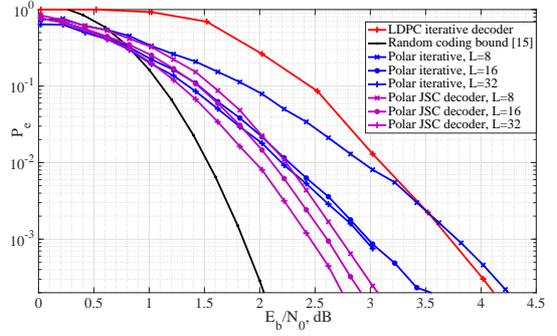}
    \caption{Probability of decoding error for $K=2$ users with code parameters $N=512, k=128$ with different list size $L$.}
    \label{fig:comparison_2users}
\end{figure}

The first experiment was conducted for $K=2$ users. The probability of decoding error~\eqref{eq:p_e} performance is shown in~\Fig{fig:comparison_2users}. Both JSC and iterative decoding schemes were tested with the list size $L=8, 16, 32$. We have selected $15$ decoding iterations for the iterative scheme. One can observe that the JSC scheme outperforms the iterative one and the LLR estimation procedure does not experience significant gain when increasing the list size. With the list size being increased the JSC also experiences higher performance gain in comparison with iterative scheme. We note, that for JSC algorithm we plotted the probability that the correct word belongs to the output list. The choice of the codeword can be done by means cyclic redundancy check (CRC) and we expect $3-5$ bit CRC to be enough. Another interesting approach is dynamically or parity-check frozen bits. We also note, that we do not need CRC for iterative decoder as the list is used only for LLR calculation.

\begin{figure}[t]
    \centering
    \includegraphics[width=0.8\columnwidth]{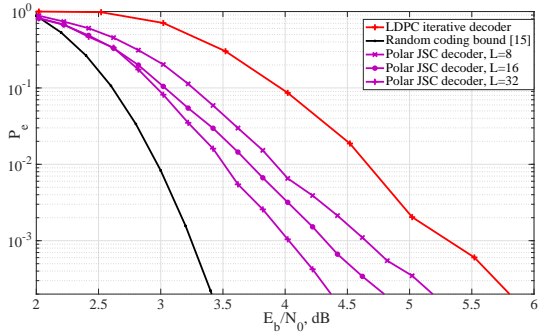}
    \caption{Probability of decoding error for $K=4$ users with code parameters $N=512, k=128$ with different list size $L$.}
    \label{fig:comparison_4users}
\end{figure}

In the second experiment we run the same schemes for $K=4$ users. We found that the result of iterative decoding is quite bad, so we used only JSC method with the same list sizes as for $K=2$ case. The results of this experiment are presented in \Fig{fig:comparison_4users}. One can easily see that JSC algorithm really improves decoding efficiency in both setups. For both cases JSC can achieve $10^{-3}$ probability of error on at least $1$ dB less energy-per-bit than PEXIT optimized LDPC. And our best-performing solution is less than $0.8$ dB apart from the random coding bound at $10^{-3}$ probability of error level. The list size also affects the performance and in case of JSC, we can see a significant performance gain when increasing the list size.
Another important practical issue is that JSC has no tunable parameters rather than iterative decoding Algorithm~\ref{alg:polar_iterative} (see \cite{Pyndiah}). We have also performed best parameters search when running iterative decoding algorithm.

\section{Conclusions and Future Work}
\vskip -0.028cm
In this paper, NOMA schemes based on polar codes are discussed. We proposed two different decoding algorithms. We have also derived a code designing procedure that optimizes polar codes for $K$-user GMAC. Then we compared our schemes with existing NOMA technique based on PEXIT optimized LDPC codes. As a result, we can conclude that JSC decoding algorithm for designed polar codes outperforms Iterative decoding procedures for both considered LDPC and polar coding schemes and becomes less that $0.8$ dB apart from the random coding bound~\citep{polyanskiy2017perspective}. We have considered single antenna AWGN model in this work and leave MIMO and fading channels for the future research.

\bibliographystyle{IEEEtran}
\bibliography{Polar.bib}

\end{document}